\documentclass[10pt,letterpaper,twocolumn]{article} 

\usepackage{ol2}
\usepackage[draft,implicit=false]{hyperref}
\usepackage{amsmath}

\begin{document}

\twocolumn[ 

\title{Modulational instability in dispersion oscillating fiber ring cavities}


\author{Matteo Conforti$^{1,*}$, Arnaud Mussot$^{1}$, Alexandre Kudlinski$^{1}$, and Stefano Trillo$^{2}$}

\address{
$^1$PhLAM/IRCICA, CNRS-Universit\'e Lille 1, UMR 8523/USR 3380, F-59655 Villeneuve d'Ascq, France\\
$^2$Dipartimento di Ingegneria, Universit\`a di Ferrara, Via Saragat 1, 44122 Ferrara, Italy\\
email: matteo.conforti@univ-lille1.fr
}

\begin{abstract}
We show that the use of a dispersion oscillating fiber in passive cavities significantly extend modulational instability to novel high-frequency bands, which also destabilize the branches of the steady response which are stable with homogeneous dispersion. 
By means of Floquet theory, we obtain exact explicit expression for the sideband gain, and a simple analytical estimate for the frequencies of maximum gain. Numerical simulations show that stable stationary trains of pulses can be excited in the cavity. 
\end{abstract}

 ] 

 \maketitle 
\baselineskip 8pt

Modulational instability (MI) refers to a process where a weak periodic perturbation of an intense continuous wave (CW) grows exponentially as a result of the interplay between dispersion and nonlinearity. The underlying mechanism is nonlinear phase matching of four-wave mixing between the CW pump and the sidebands, which requires, for the scalar process in a homogeneous optical fiber, anomalous group-velocity dispersion (GVD).
In the normal GVD regime, MI can occur in detuned cavities, thanks to the constructive interference between the input pulse and the attenuated pulse that returns after a round-trip \cite{haelterman92,haelterman92oc,coen97,coen01,leo13}. Alternatively it can arise in systems with built-in periodic dispersion \cite{kelly92,matera93,bronski96,smith96,Abdullaev96,abdullaev99}, among which dispersion oscillating fibers (DOFs) have recently attracted a lot of attention \cite{armaroli12,droques12,droques13,finot13}. In this case, phase matching relies on the additional momentum carried by the periodic dispersion grating (quasi-phase-matching). 

In this letter, we investigate the combination of the two mechnisms, namely a passive fiber cavity with built-in DOF, showing that, despite the simplicity of implementation, this type of structures exhibit novel interesting features. The additional periodicity introduced by the DOF not only extends MI to the branches of the bistable response which are stable when GVD is homogeneous, but also induces higher-frequency MI branches which could be conveniently exploited for pulse train generation at high repetition rate.
Moreover, MI can rise in the monostable regime of the cavity  even in normal dispersion. Importantly, after the stage of exponential growth of weak sidebands, DOF-induced MI can generate stable pulse trains with large contrast ratio at frequencies which are comparatively higher than those of the homogeneous GVD case. 
Theoretical analysis based on Floquet theory is supported by numerical simulations of the Lugiato-Lefever Equation (LLE) \cite{haelterman92, coen13,LL87} and the cavity map. 

We consider a fiber ring modeled by cavity boundary conditions for the $n-$th round-trip, coupled to Nonlinear Schr\"odinger Equation (NLSE), that rules the propagation of the intracavity field $u_n$ along the ring:
\begin{align} 
&u_{n+1}(z=0,t)=\theta u_{in}(t) +	\rho e^{-i\delta}u_n(z=1,t), \label{map1}\\\
&i\frac{\partial u_n}{\partial z}-\frac{\beta(z)}{2}\frac{\partial^2 u_n}{\partial t^2}+|u_n|^2u_n=0 \label{map2},
\end{align}
where subscript $n$ indicates $n-$th circulation, $\rho, \theta$ ($\rho^2+\theta^2=1$) are the reflection and transmission coefficients, $\delta$ is the cavity detuning and we define dimensionless units as follows: $z=Z/L$, $t=T/T_0$, $u(z,t)=E(Z,T)\sqrt{\gamma L}$, $T_0=\sqrt{|k''|L}$, $\beta(z)=k''(z)/|k''|$ where $L$ is the cavity length, $k''=d^2k/d\omega^2$ the average second order dispersion and $\gamma$ the fiber nonlinearity.  Quantities in capital letters $Z$, $T$, $E$ denotes real world distance, retarded time in the frame traveling at group velocity, and intracavity electric field envelope, respectively.
In a mean-field approach, the map (\ref{map1}-\ref{map2}) can be averaged to give the following LLE \cite{haelterman92,haelterman92oc}
\begin{equation}\label{lle}
i\frac{\partial u }{\partial z}-\frac{\beta(z)}{2}\frac{\partial^2 u}{\partial t^2}+|u|^2u= \left(\delta-i\alpha \right)u + iS,
\end{equation}
where we drop the subscript for the field and set $S=\sqrt{P}=\theta u_{in}$, where $u_{in}=\sqrt{\gamma L}E_{in}$ is the normalized input external field, and $\alpha=1-\rho \approx\theta^2/2$ describes cavity losses (generally dominated by output coupling). While Eq. (\ref{lle}) is equivalent to unfold the cavity, its round-trip periodicity constrains $z$ to be accessible at integer values. Consistently we consider a normalized DOF period $\Lambda=1/N$, $N=1,2,\ldots$, where $N$ represents the number of periods of the DOF over a single round-trip. 

It is well known that Eq. (\ref{lle}) shows bistability with two coexisting stable branches of CW solutions $u=u_0$, whenever $\delta^2>3\alpha^2$ \cite{haelterman92oc,LL87}. This follows from the inverse steady-state response $P=P(P_u)$
\begin{equation} \label{LLEstaz}
P=P_u[(P_u-\delta)^2+\alpha^2],
\end{equation}
where $P=|S|^2$ and and $P_u=|u_0|^2$ are driving and intracavity powers, respectively. The values $P_u=\left(2\delta\pm\sqrt{\delta^2-3\alpha^2}\right)/2$ mark the bistable knees \cite{haelterman92oc,LL87}.

Cavity steady-states can destabilize according to MI, entailing the exponential growth $\propto \exp[g(\omega)z]$ of periodic modulations with proper frequency $\omega$. 
MI can be characterized by means of the linear stability analysis of the steady solution $u_0$. We consider the evolution of a perturbed solution $u(z,t)=u_0+\eta(z,t)$, with $|\eta|\ll|u_0|$ and express the perturbation as the combination of two symmetric sidebands $\eta(z,t)=\varepsilon_s(z)\exp[i\omega t]+\varepsilon_a(z)\exp[-i\omega t]$. We obtain a linear ODE system for the perturbations $d\varepsilon/dz=M(z)\varepsilon$, $\varepsilon=[\varepsilon_s,\varepsilon_a^*]^T$, where the matrix $M=M(z)$, reads:
\begin{equation}\label{M}
M=
\left[
\begin{array}{ c c}
i\Omega^2(z)-\alpha & i u_0^2\\
-i u_0^{2*} & -i\Omega^2(z)-\alpha
\end{array}
\right],
\end{equation}
with $\Omega^2(z) \equiv \frac{\beta(z)}{2}\omega^2+2 P_u-\delta$.
Whenever $\beta$ is constant, we recover the MI gain in closed form as \cite{haelterman92,haelterman92oc}
\begin{equation}
g(\omega)=-\alpha+\sqrt{4 P_u \delta_\omega -\delta_\omega^2 -3 P_u^2}, \;\;\delta_\omega=\delta-\frac{\beta}{2}\omega^2,
\end{equation}
which yields the most unstable frequency and its gain,
\begin{equation}\label{gain}
\omega_{max}=\sqrt{\frac{2}{\beta}(\delta-2 P_u)},\;\; g(\omega_{max})= P_u -\alpha.
\end{equation}
If we consider a periodic dispersion $\beta(z)=\beta(z+\Lambda)$, we can analyze MI by applying Floquet theory \cite{bronski96,nayfeh}. The stability depends on the so-called Floquet multipliers or characteristic exponents. These are obtained by constructing the $2\times 2$ fundamental matrix solution evaluated at $z=\Lambda$,  $U=[y_1(\Lambda),y_2(\Lambda)]$, whose columns are the solutions $y_{1,2}(z)$ to $d\varepsilon/dz=M(z)\varepsilon$, for the two initial values $\varepsilon(0)=[1,0]^T,[0,1]^T$. MI occurs when one of the eigenvalues $\lambda$ of $U$ is such that $|\lambda|>1$. In this case the (amplitude) growth rate of the instability is given by the characteristic exponent $g(\omega)=\ln|\lambda|/\Lambda$.

\begin{figure}
	\centering
		\includegraphics[width=0.23 \textwidth]{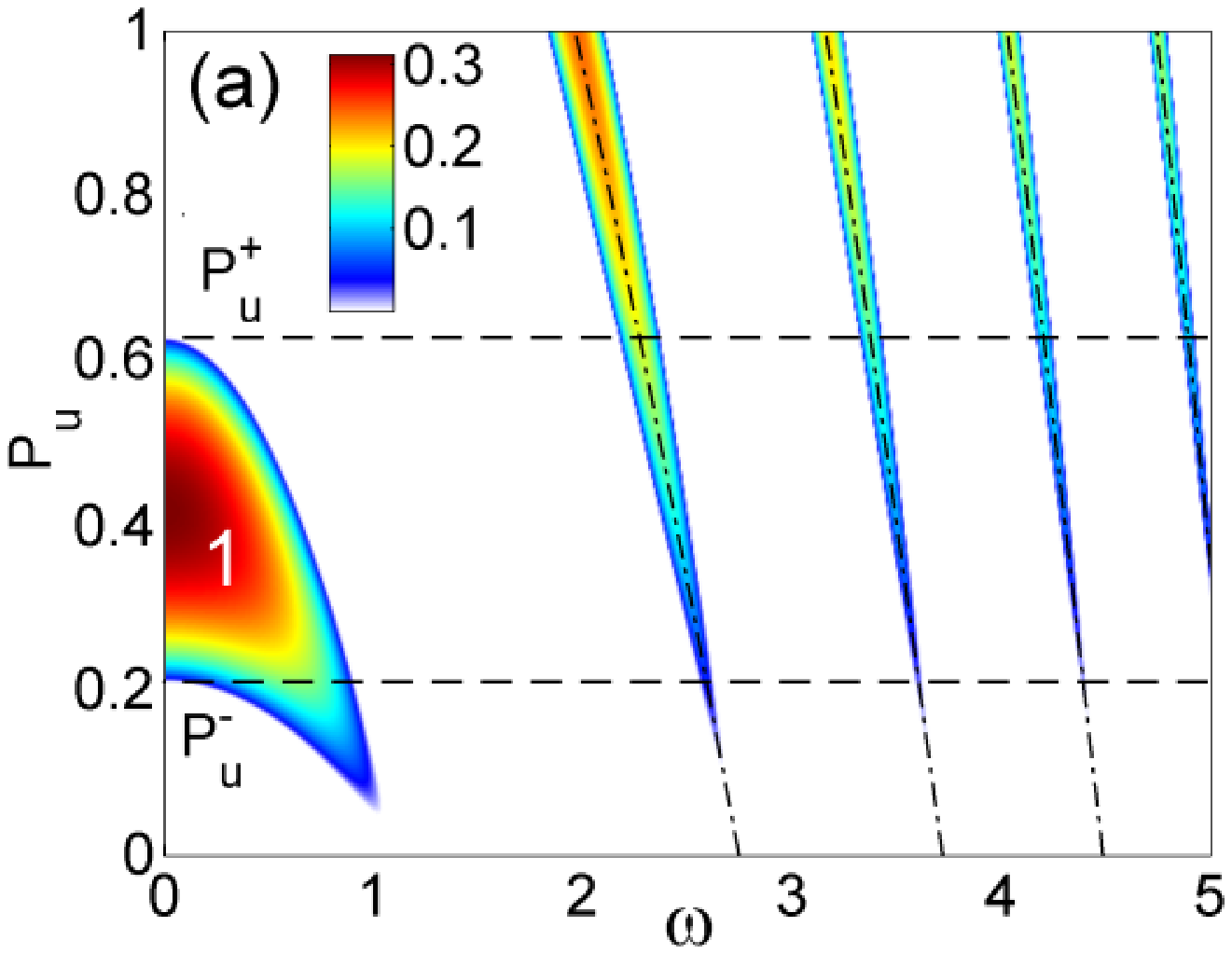}
\includegraphics[width=0.23 \textwidth]{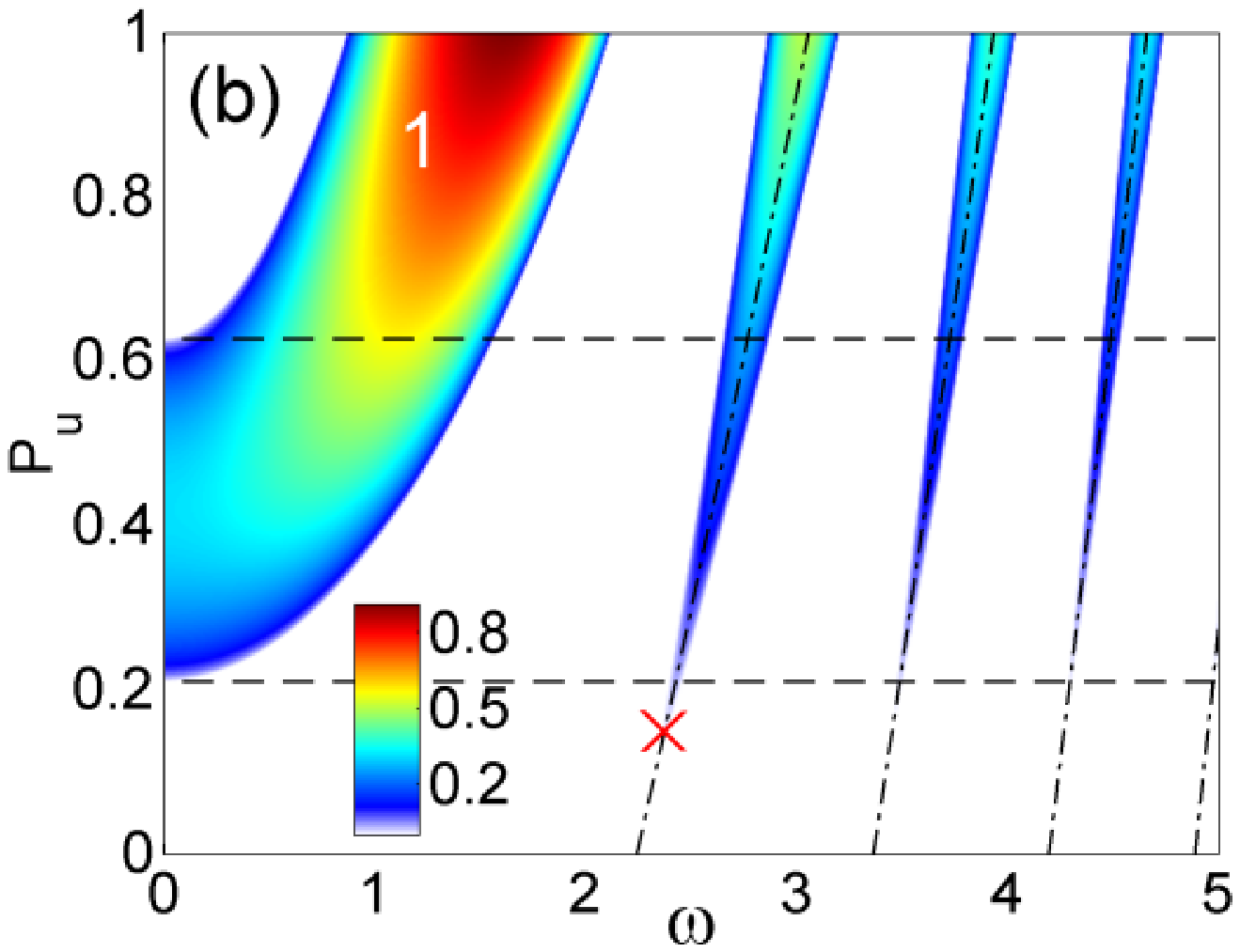}
\includegraphics[width=0.23 \textwidth]{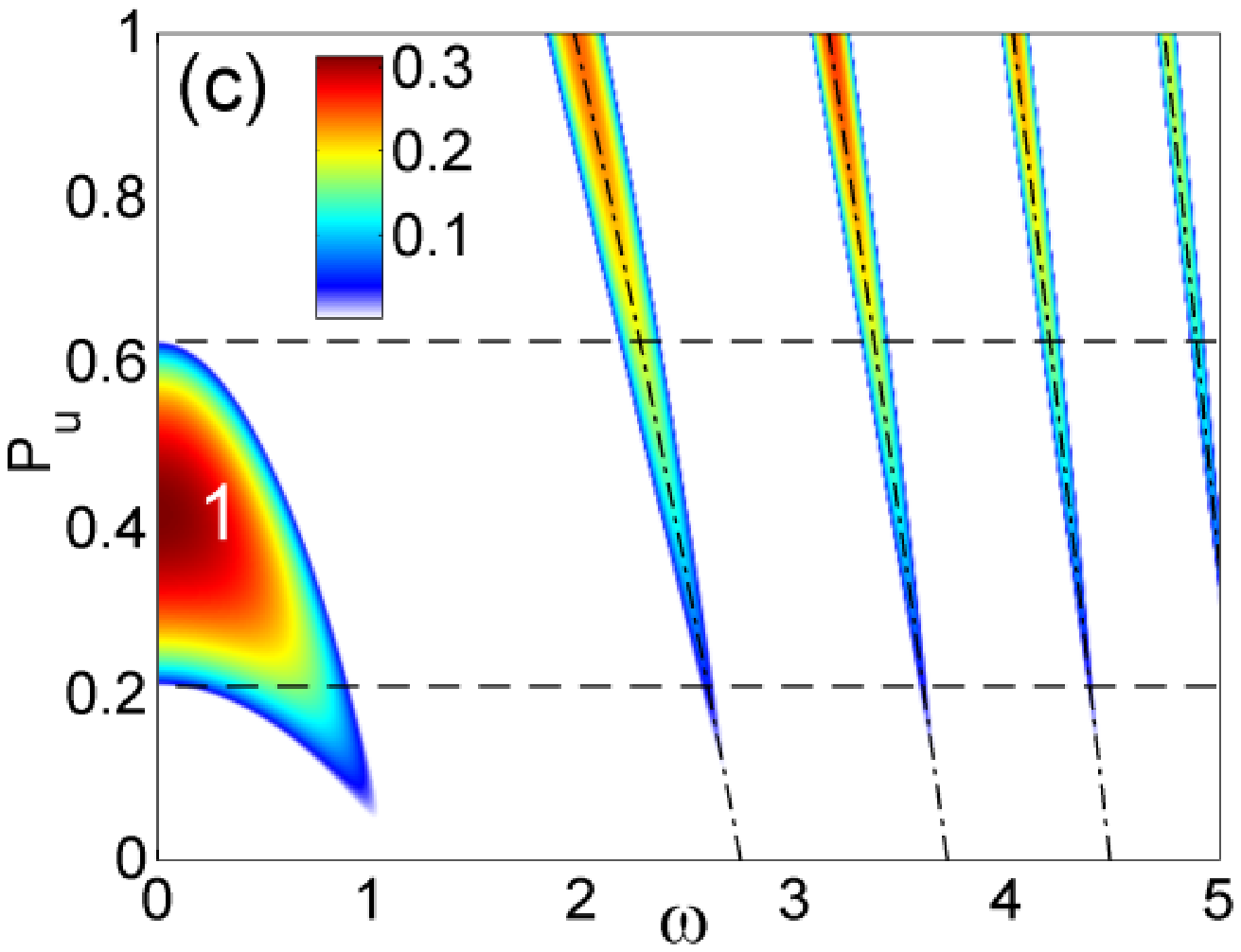}
\includegraphics[width=0.23 \textwidth]{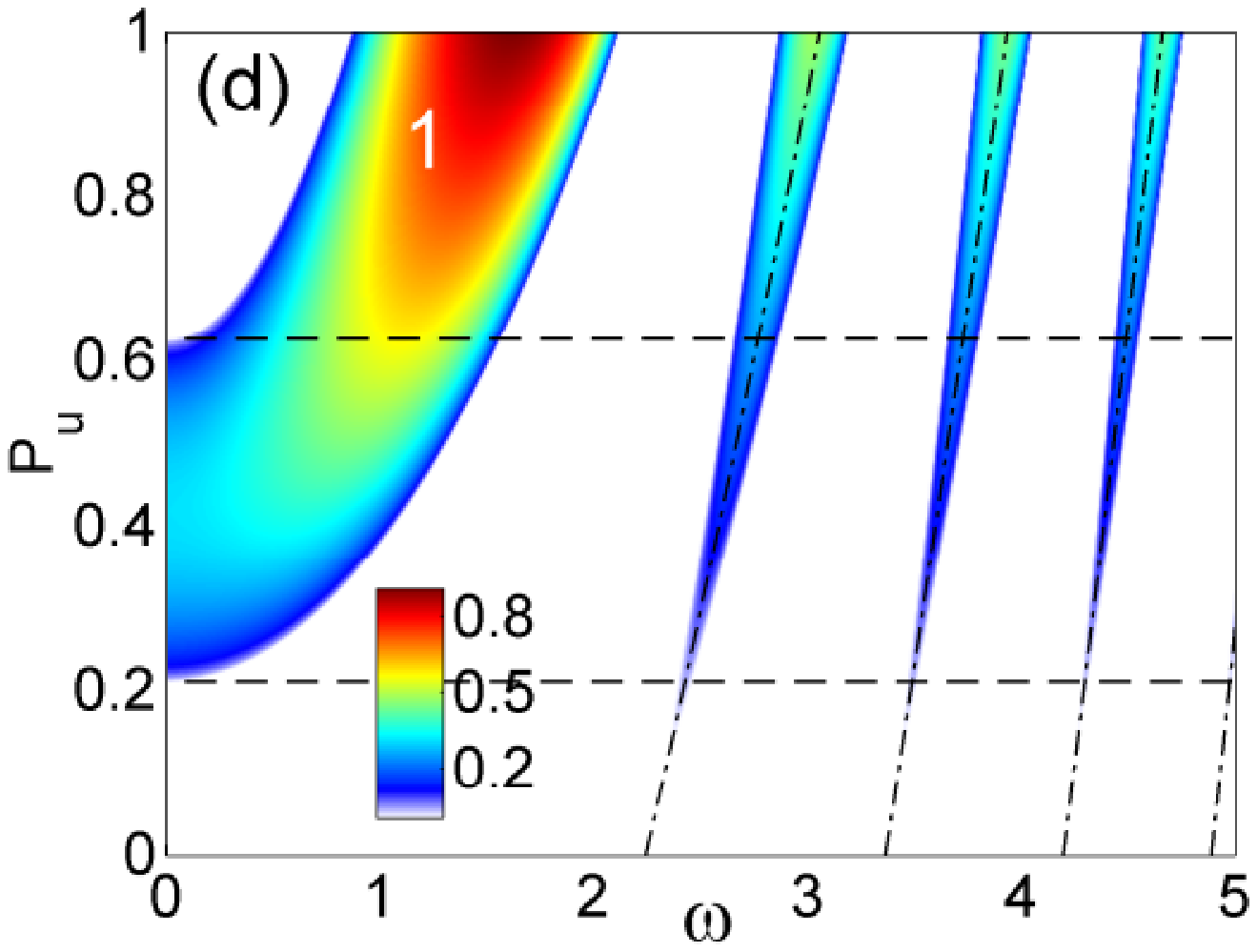}
\includegraphics[width=0.4 \textwidth]{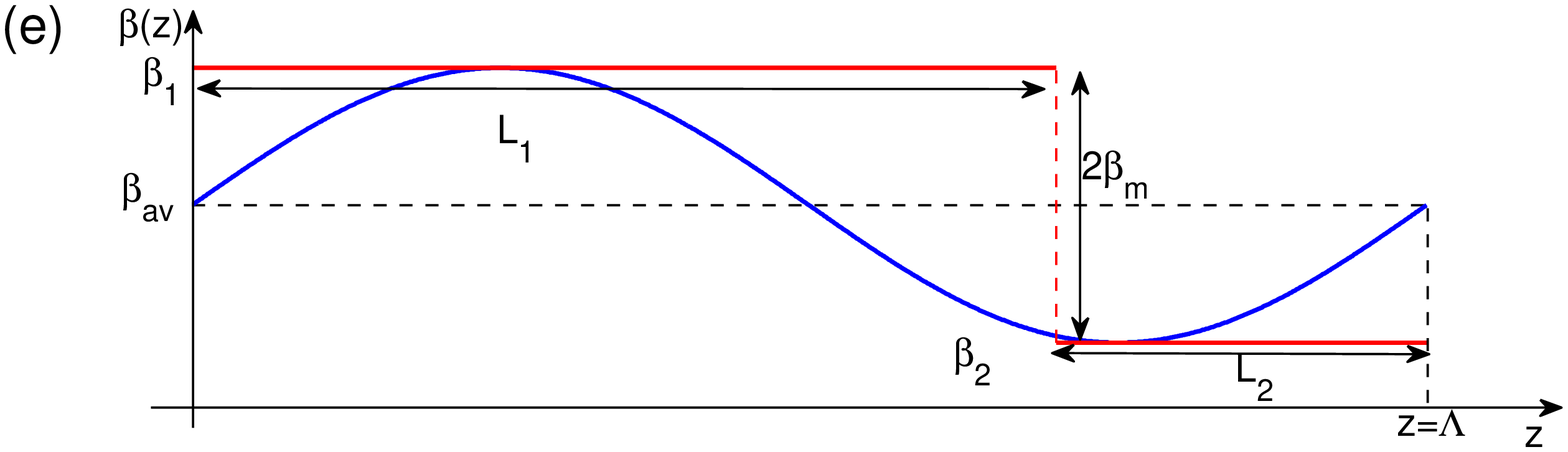}
	\caption{Level plot of MI gain in the plane ($\omega,P_u$) with normal (a,c) and anomalous (b,d) average GVD. (a,b) Sinusoidal dispersion $\beta(z)=\pm 1+\sin(2\pi z)$; (c,d) Piecewise constant dispersion $L_1=L_2=1/2$, $\beta_1=\pm 1+0.8$, $\beta_2=\pm 1-0.8$. Dash-dotted black curves, MI peak gain frequencies from Eq. (\ref{wmax}). Horizontal dashed lines stand for $P_u^{\pm}$ ($P_u^- < P < P_u^+$ correspond to the negative slope branch of the bistable response). Number 1 denotes the MI branch of the homogeneous cavity. Parameters: $\delta=\pi/5$,  $\alpha=\theta^2/2=0.05$. (e) Sketch of periodic dispersion profiles over one period: sinusoidal (blue curve) and piecewise constant (red curve). Average dispersion $\beta_{av}=(\beta_1+\beta_2)/2$, modulation depth $\beta_m=(\beta_1-\beta_2)/2$.}
	\label{fig1}
\end{figure}

For a generic profile of GVD (e.g. sinusoidal) the problem cannot be solved analytically and some approximated results can be found by exploiting the method of averaging \cite{armaroli12}; otherwise we need to resort to numerical methods. However, if $\beta(z)$ is piecewise constant, i.e. we have two sections of length $L_{1,2}$ ($L_1+L_2=\Lambda$) with dispersion $\beta=\beta_{1,2}$ [Fig. \ref{fig1}(e)], we can find an exact solution. 
In this case, the eigenvalues of $U$, i.e. the characteristic exponents, are given by
\begin{equation}\label{lambda}
\lambda_{1,2}=\frac{\Delta}{2}\pm\sqrt{\frac{\Delta^2}{4}-W},
\end{equation}
where $\Delta=y_{11}+y_{22}$ and the Wronskian $W=y_{11}y_{22}-y_{12}y_{21}$ is easily calculated as $W=e^{-2\alpha \Lambda}$ [$y_{mn}$ is the $n-$th component of the solution $y_m(\Lambda)$].
We have MI if $|\Delta|>(1+W)$, with gain $g(\omega)=\ln(\max|\lambda_{1,2}|)/\Lambda$, where $\Delta$ is the Floquet discriminant
\begin{equation}\label{delta}
\Delta=e^{-\alpha L}[2\cos(k_1L_1)\cos(k_2L_2)-\sigma \sin(k_1L_1)\sin(k_2L_2)],
\end{equation}
where $\sigma=[\beta_1\beta_2\omega^4+2( P_u -\delta)(\beta_1+\beta_2)\omega^2+4(3 P_u -\delta)( P_u -\delta)]/(2k_1k_2)$, and $k_{1,2}=\sqrt{(\beta_{1,2}\omega^2/2+2P_u-\delta)^2-P_u^2}$.
%
From Eq. (\ref{delta}) it is possible to extract some useful analytical estimates. In particular, in the limit of small modulation depth ($\beta_1\approx\beta_2$, $\beta_m\approx 0$),
we can easily show that $\sigma\approx 2$.  Moreover, if we consider $L_1=L_2=\Lambda/2$, it is possible to obtain a very simple expression of the frequencies of the peak MI gain. 
This is given by the relation $(\beta_{av}\omega^2/2+2P_u-\delta)^2-P_u^2=(m\pi  /\Lambda)^2$, $m=1,2,\ldots$, where one can recognize the condition of parametric resonance, i.e. the natural frequency of the unperturbed oscillator (\ref{M}) ($k_{av}$) is a multiple of half the forcing frequency ($\pi/\Lambda$) \cite{armaroli12}. We get explicitly

%
\begin{equation}\label{wmax}
\omega_{max}=\sqrt{\pm\left[\frac{2}{\beta_{av}}\sqrt{\left(\frac{m\pi}{\Lambda}\right)^2+P_u^2}\right]+\frac{2}{\beta_{av}}\left(\delta - 2P_u \right)},
\end{equation}
%
where the contribution of the DOF appears in the term in square brackets [cfr. Eqs. (\ref{wmax}) and (\ref{gain})].

Figure \ref{fig1} shows relevant examples of MI gain domains for a bistable response ($\delta=\pi/5$,  $\alpha=0.05$), in both the normal and anomalous GVD regimes, with DOF period $\Lambda=1$ (henceforth, we show results for this case that entails a GVD period exactly equal to the cavity length; as a general remark, shorter periods give qualitatively similar results, with MI tongues shifting towards higher frequencies). Figure \ref{fig1}(a) refers to a sinusoidal GVD with average value $\beta_{av}=1$ ({\em normal} GVD) and modulation depth $\beta_m=1$.  
The low frequency branch (labeled 1) in Fig. \ref{fig1}(a) is characteristic of the average value of GVD. Indeed, similarly to the homogeneous case \cite{haelterman92}, this branch entails MI of the lower branch of the steady response ($0 \le P_u \le P_u^-$), which extends over the negative slope of the response ($P_u^- < P_u < P_u^+$), where also CW perturbations ($\omega=0$) are unstable. 
However, additional birth of new high-frequency MI tongues can be clearly seen. Importantly, these narrowband branches of MI are responsible for destabilizing the upper branch of the response ($P_u^+ \le P_u$).
Figure \ref{fig1}(b) refers to the same parameters (bistable response) as in Fig. \ref{fig1}(a), but {\em anomalous} average GVD, $\beta_{av}=-1$. 
In this regime the cavity imposes a power threshold for MI to develop (even when $\alpha=0$) \cite{haelterman92, haelterman92oc},
resulting in the opposite situation (stable lower branch, unstable upper branch), as shown by the low-frequency tongue (label 1) in Fig. \ref{fig1}(b), reminiscent of the homogeneous case. The oscillating dispersion induces additional MI tongues to appear in this case too. Remarkably, in this case they destabilize the lower branch of the steady state response. The MI tongues have different power thresholds (that exists, due to the losses), with the lowest one arising, in this case, for the first tongue.

Figure \ref{fig1}(c) shows the MI gain for a piecewise constant dispersion [$\beta_1=0.2$, $\beta_2=1.2$, $L_1=L_2=1/2$], where we can exploit the analytical expressions (\ref{lambda},\ref{delta}). The results are very similar to the case of sinusoidal variation (provided we consider the same period and a 50-50 duty cycle). More precisely, MI tongues appear at the same frequencies, though we observe a slight variation of the gain. 
The same analogy holds also in the anomalous dispersion regime [compare Fig. \ref{fig1}(b) and Fig. \ref{fig1}(d)], showing that the analytical formulas [Eqs. (\ref{lambda}, \ref{delta})] give insight for general dispersion profiles.


In all the examples reported in Fig. 1, we note that the peak gain position analytically predicted from Eq. (\ref{wmax}) accurately reproduces the numerical results, despite we are considering either relatively strong dispersion modulation or sinusoidal dispersion profiles.

\begin{figure}
	\centering
		\includegraphics[width=0.23 \textwidth]{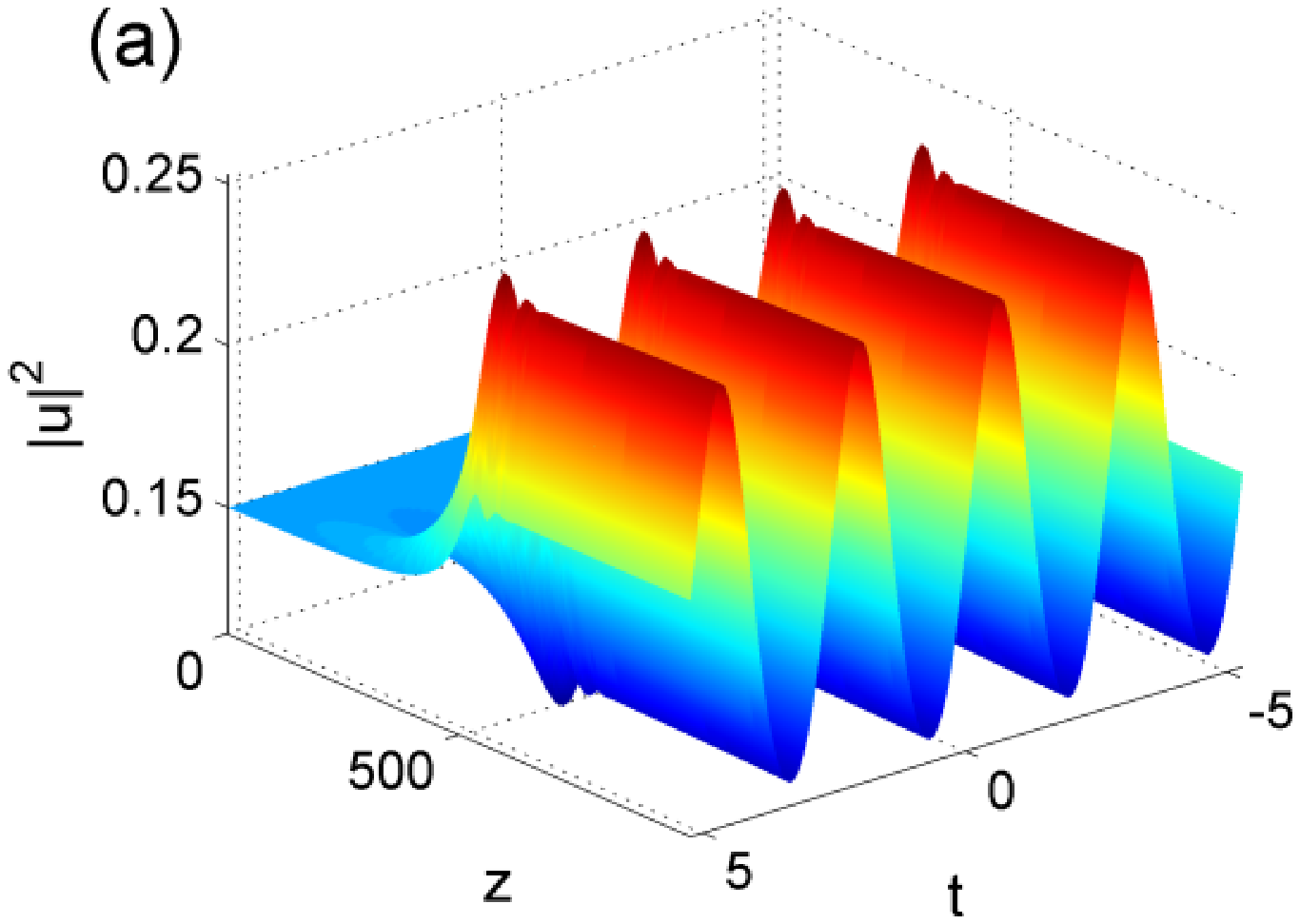}
\includegraphics[width=0.23 \textwidth]{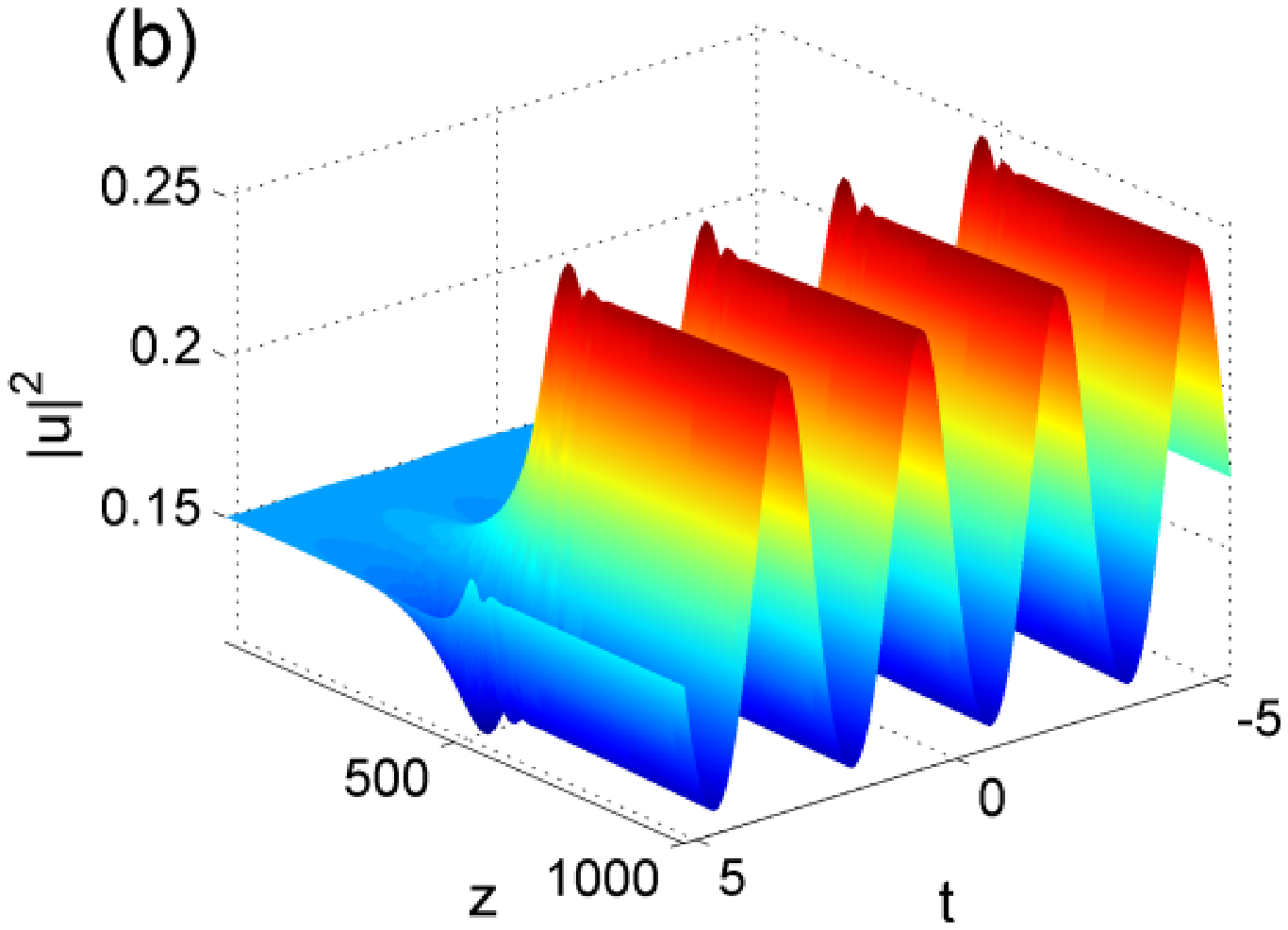}
\includegraphics[width=0.23 \textwidth]{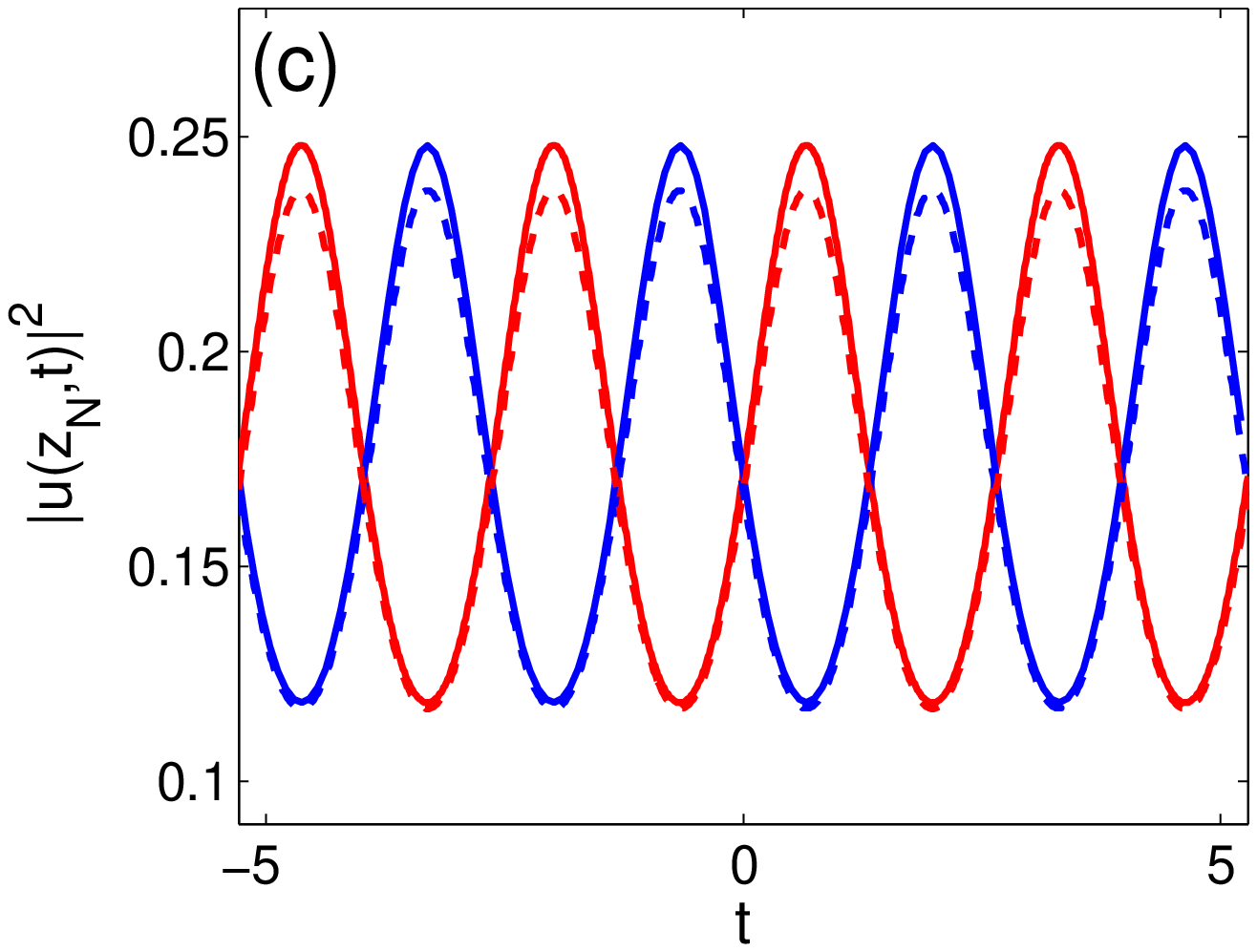}
\includegraphics[width=0.23 \textwidth]{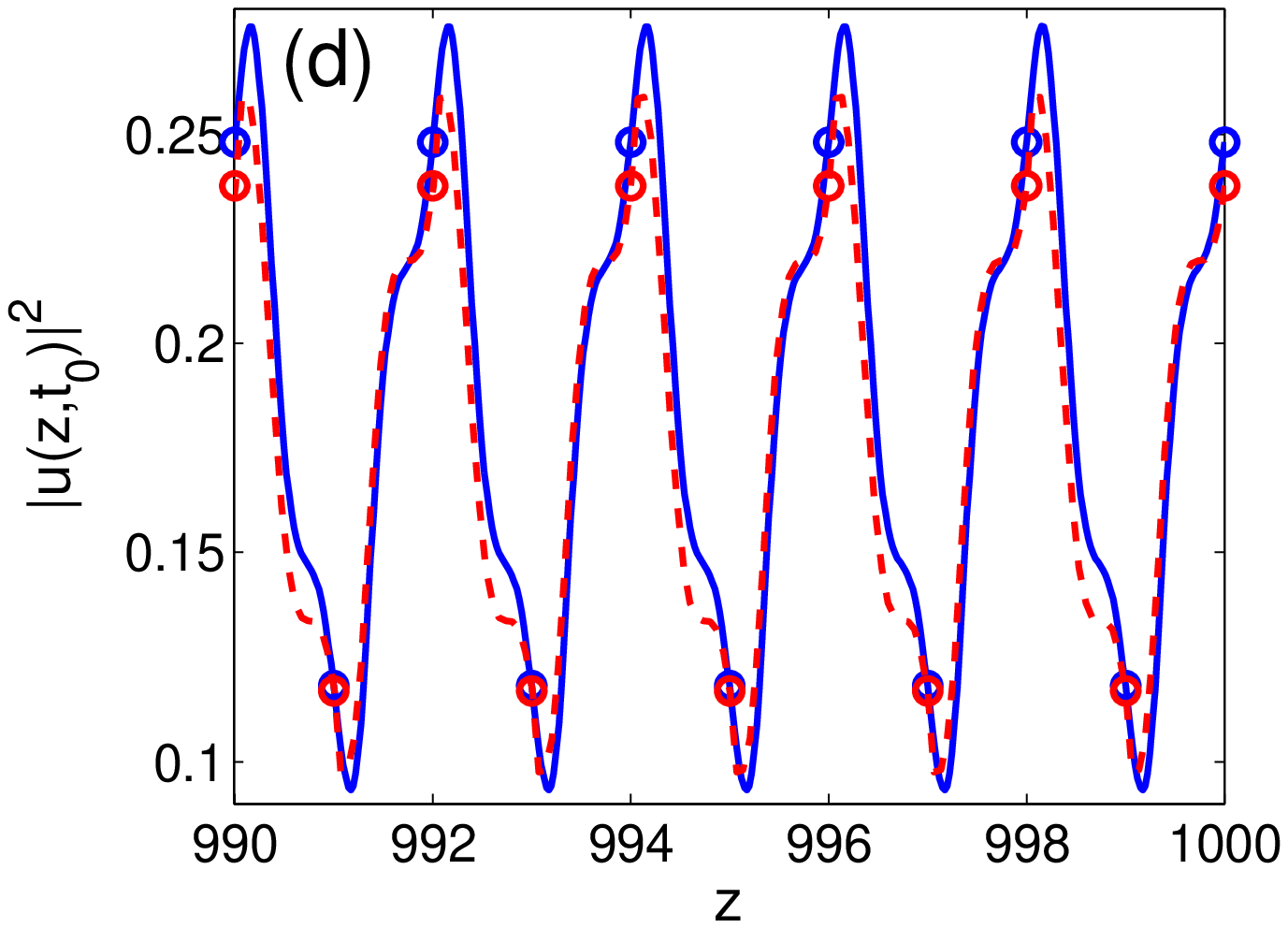}
	\caption{(a,b) Intracavity power $|u|^2$ at integer $z$ (round-trips) from numerical solution of LLE: (a) even round-trips; (b) odd round-trips. (c) Stationary pulse trains at even (blue curves) and odd (red curves) round-trips. Dashed curves are numerical solution of the map [Eqs. (\ref{map1}-\ref{map2})]. (d) Power $|u(z,t_0)|^2$ evaluated at $t=t_0$ corresponding to a maximum of the pulse train; Solid blue curve, LLE; dashed red curve, map. Dots correspond to observable field at the output coupler. 
Parameters: $\beta(z)=-1+\sin(2\pi z/\Lambda)$ (average anomalous GVD), $\Lambda=1$, $\delta=\pi/5$,  $\alpha=\theta^2/2=0.05$, $P_u=0.15$ (lower branch of steady state), which yield $\omega_{max}=2.37$ [red cross in Fig. \ref{fig1}(b)] with gain $g(\omega_{max})=0.01$.}
	\label{fig2}.
\end{figure}
We have verified the result of linear stability analysis through extensive simulations of LLE (\ref{lle}). In a relatively large range of parameters, MI induced by DOF generates, beyond the stage of exponential growth, stable pulse trains with large and fixed contrast ratio which survives indefinitely.
Figure \ref{fig2}(a,b) shows the development of a pulse train from a weak seed at the peak MI gain frequency of the first tongue [see red cross in Fig. \ref{fig1}(b)]. 
Here we considered anomalous 
average GVD and a sinusoidal DOF with an input power just above threshold for the MI over the first tongue. 
In contrast with what reported for homogeneous fiber cavities \cite{haelterman92,haelterman92oc}, here the cavity stabilizes over a period-2 (P2) attractor \cite{ikeda82}, such that the pulse train is reproduced identical after two round-trips.
Between two successive round-trips the pulse train turns out to be simply out-of-phase \cite{haelterman92p2}, which makes this P2-state peculiar and different from those developing from self-pulsing instabilities\cite{ikeda82}. 
Out of phase pulse train attractors have already pointed out to exist in the LLE \cite{LL87,haelterman92p2}, being ultimately linked to the symmetry of the periodic eigenstates of the NLSE \cite{tw91}.
This behavior is well described in Figs. \ref{fig2}(a) and (b), that show the intracavity field sampled at even (a) and odd (b) round-trips. A physical explanation of this phenomenon is that the MI sidebands generated by means of DOF are not in-phase with the pump \cite{droques13}, and acquires a phase shift at each period, that corresponds in our case to the fiber length. This evolving phase between the pump and the sidebands, makes the field fluctuate over the length-scale of the cavity, as depicted in Fig. \ref{fig2}(d). This fact can shed some doubts on the validity of the LLE model, which is derived assuming that the field evolves slowly over several round-trips. However direct simulations of the map [Eqs. (\ref{map1}-\ref{map2})] confirm the validity of the analysis based on LLE, as shown 
in Figs. \ref{fig2}(c-d). The extended validity of LLE can be understood if we consider that the field is not evolving freely: in fact the power of the spectrum is evolving slowly, and only the spectral phase can change rapidly. Moreover the cavity boundary conditions modifies slightly and slowly the field (which is the true requirement for applicability of mean-field model), whereas the rapid variations during each round-trip are given by the propagation ruled by NLSE.

\begin{figure}
	\centering
		\includegraphics[width=0.23 \textwidth]{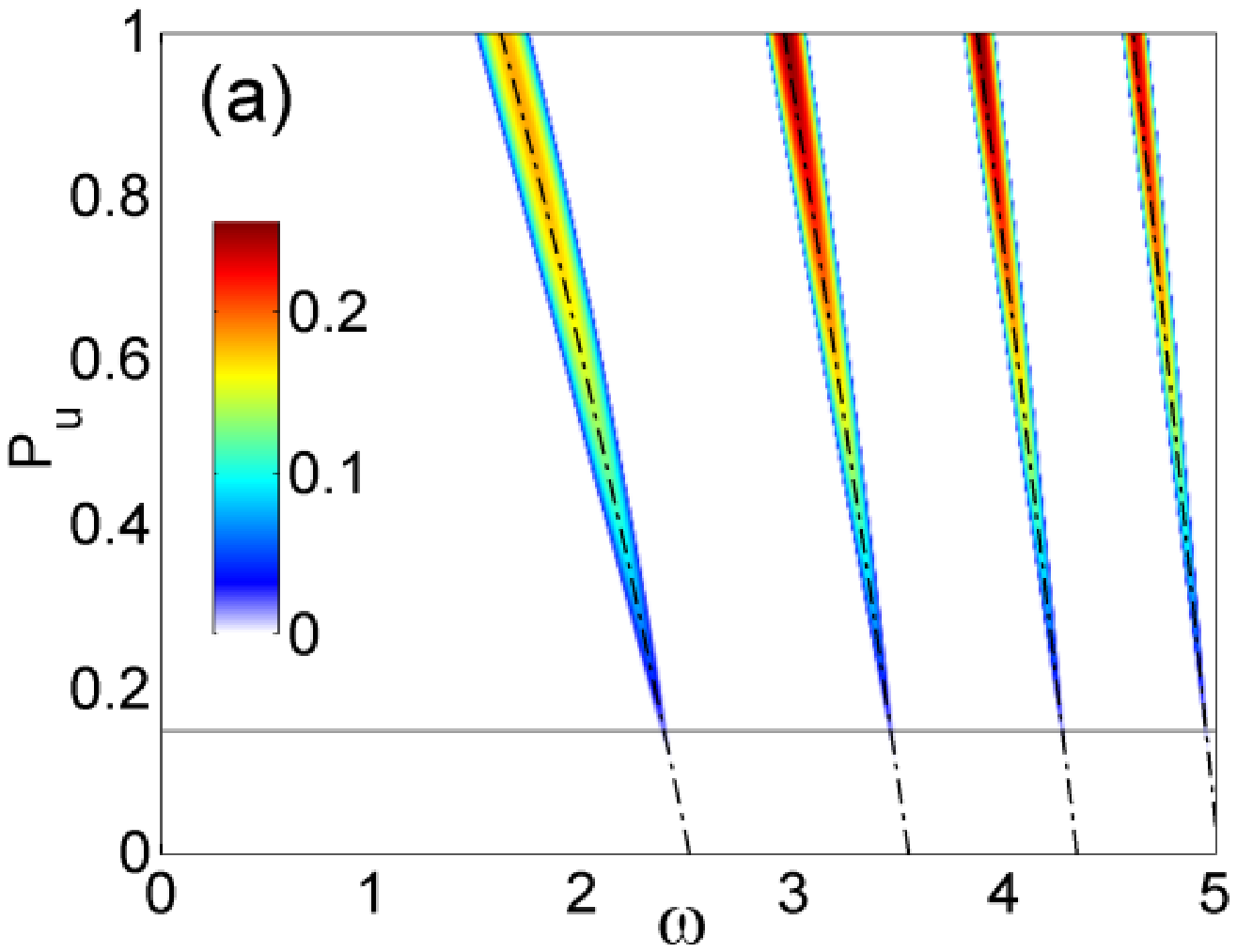}
\includegraphics[width=0.23 \textwidth]{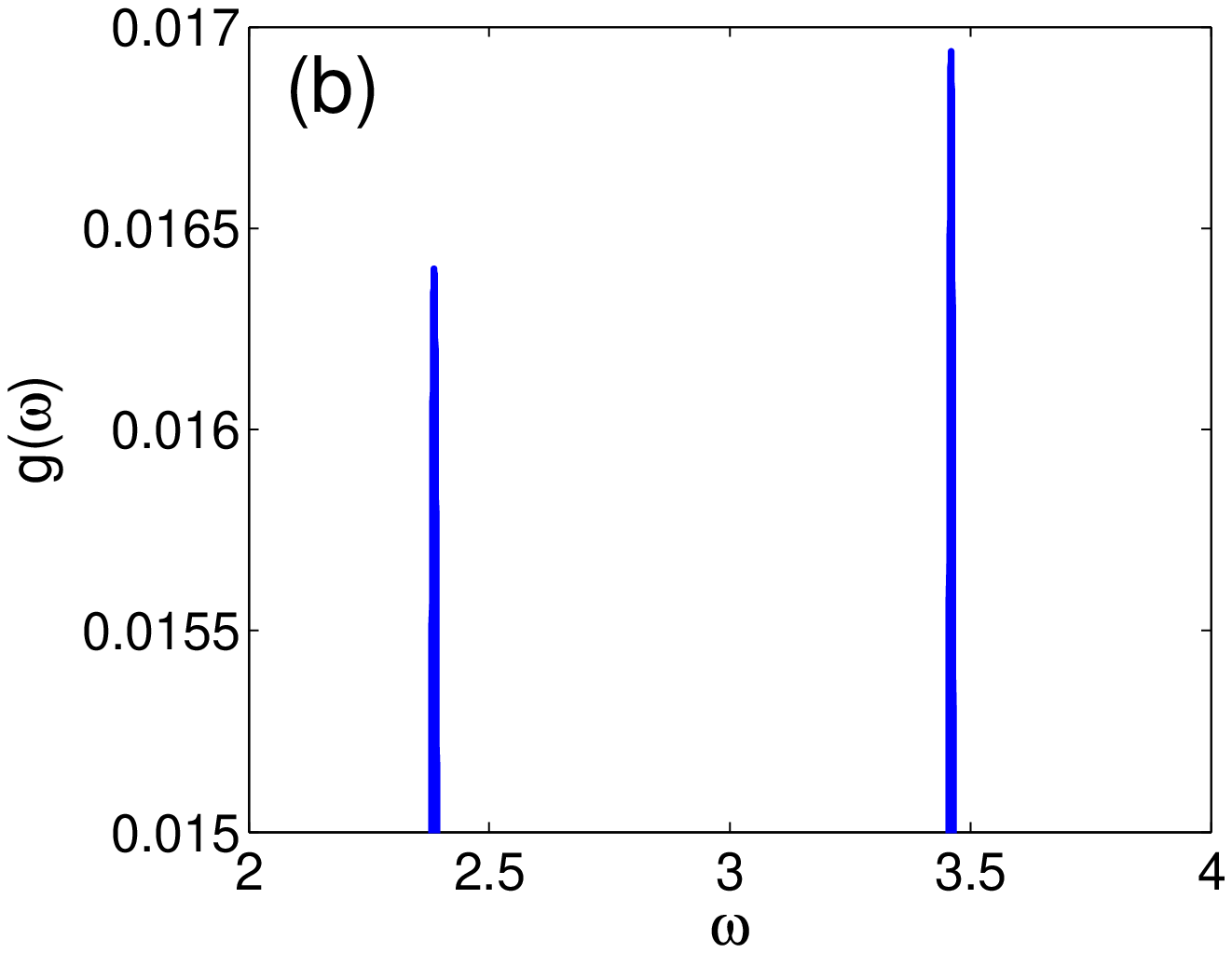}
\includegraphics[width=0.23 \textwidth]{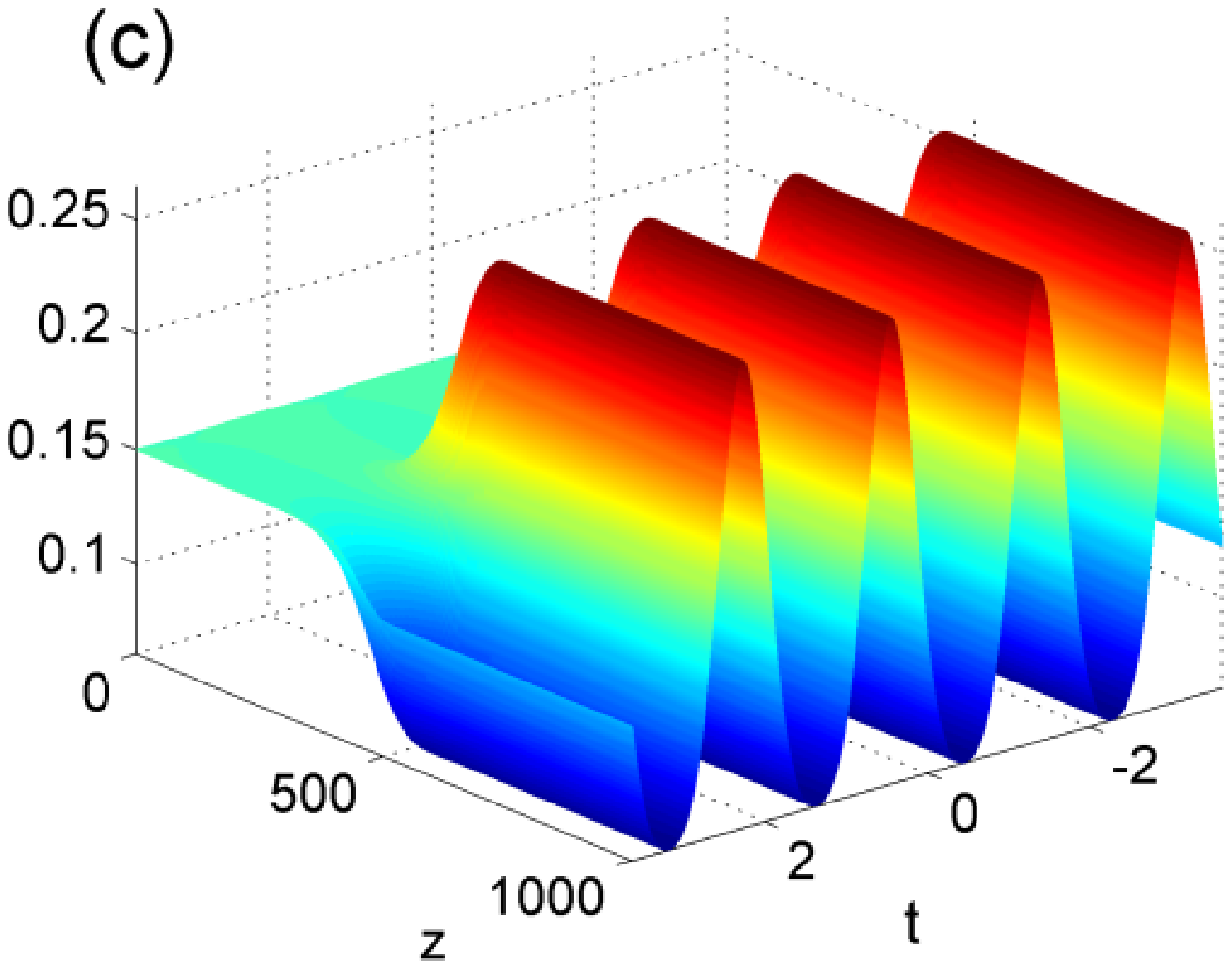}
\includegraphics[width=0.23 \textwidth]{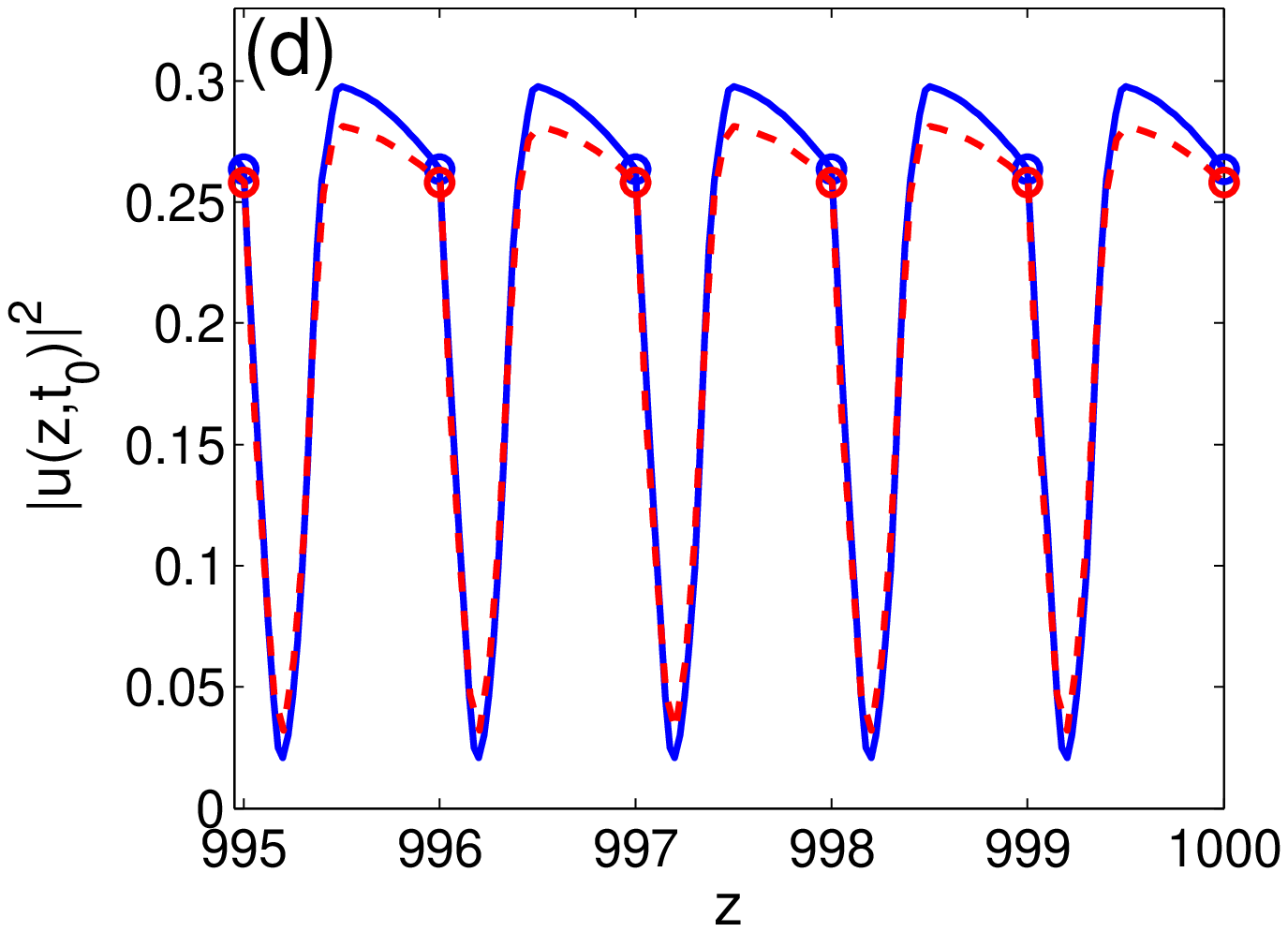}
	\caption{(a) Level plot of MI gain in  monostable regime with piecewise-constant average normal GVD. Dash-dotted black curves, MI peak gain frequencies from Eq. (\ref{wmax}). (b) Section at $P_u=0.15$  [horizontal black line in panel (a)], showing the first two MI tongues. (c) Intracavity field power at each roundrip $z=1,2,\ldots$, calculated from numerical solution of LLE Eq. (\ref{lle}). (d) Intracavity field power evolution evaluated at the time $t_0$, corresponding to a maximum of the pulse train; Solid blue curve, LLE; dashed red curve, map. Dots correspond to observable field at the output coupler. Parameters: $\beta_{1,2}=1\pm 0.9$, $L_1=L_2=\Lambda/2=0.5$, $\delta=0$,  $\alpha=\theta^2/2=0.05$, $P_u=0.15$.}
	\label{fig3}
\end{figure}

DOF-induced MI can appear even in the monostable and normal dispersion regime of the cavity, where standard cavity MI is forbidden. Figure \ref{fig3}(a) shows the MI gain for a piecewise constant dispersion  $\beta_{1,2}=1\pm 0.9$: we can see that the homogeneous cavity MI band is absent, whereas DOF induced tongues are still present. Interestingly enough, for this value of modulation, the second tongue has higher gain than the first one [see Fig. \ref{fig3}(b)], so one can envisage the cavity dynamics to be driven by the higher frequency band. Numerical simulation of the LLE (\ref{lle}) shows again attractive behavior towards a periodic solution that represents a stable pulse train. However, in this case, we observe a period-1 solution [see Fig. \ref{fig3}(c)]. The intracavity field displayed in Fig. \ref{fig3}(d) exhibits fast oscillations in this case too, with the round-trip periodicity. Good agreement is found between the LLE [Eq. (\ref{lle})] and the map [Eqs. (\ref{map1}-\ref{map2})].

Let us consider physical parameters for a typical fiber ring cavity \cite{leo13}. We can take a fiber of length $L=100$ m, average dispersion $k''=1$ ps$^2$/km and nonlinear coefficient $\gamma=2.5$ W$^{-1}$km$^{-1}$ . Characteristic power and time units are $P_c=(\gamma L)^{-1}=4$ W, $T_c=\sqrt{k'' L}=316$ fs.  The example reported in Fig. \ref{fig2} corresponds to input power of $1.4$ W, with MI frequency of $1.6$ THz.

In summary, we have shown that a dispersion oscillating fiber ring cavity develops new forms of MI. By exploiting Floquet theory we calculated numerically MI gain for different dispersion profiles. Quite remarkably, we were able to find analytical expressions for the relevant case of piecewise constant dispersion. Moreover, this analytical expression can give insight also for other dispersion profiles (e.g. sinusoidal) of the same periodicity. We numerically showed that the nonlinear development of MI give birth to stable pulse trains pumped both in the normal and anomalous dispersion regimes. 
These pulse trains can exhibit periodicity over multiple round-trips, depending on the MI tongue that excites them. 
We have also observed that, despite the field does not change slowly over the scale of several round-trips, the evolution is correctly captured by the LLE, witnessing once more the power of this simple model for the description of even complex physical set-up \cite{coen13}. 

Funding from French National Research Agency (grant TOPWAVE) and Italian Ministry of Research (grant PRIN 2012BFNWZ2) is gratefully acknowledged.


\clearpage

\section*{Informational Fifth Page}

\end{document}